\documentclass[letterpaper, 10pt]{article}
  
  \usepackage{geometry}
\usepackage{times}			
\usepackage{acronym}
\usepackage{cite}
\usepackage{graphicx}
\usepackage{amsmath}
\usepackage{notoccite}
\usepackage{xcolor}
\usepackage{amsfonts}
\usepackage{amssymb}
\usepackage{dsfont}			

\acrodef{amfm}[AM-FM]{Amplitude-and Frequency-Modulated}
\acrodef{if}[IF]{Instantaneous Frequency}
\acrodef{ia}[IA]{Instantaneous Amplitude}
\acrodef{mcs}[MCS]{Multicomponent Signal}
\acrodef{stft}[STFT]{Short-Time Fourier Transform}
\acrodef{tfr}[TFR]{Time-Frequency Representation}
\acrodef{tf}[TF]{Time-Frequency}
\acrodef{rd}[RD]{Ridge Detector}
\acrodef{mrf}[MRF]{Markov Random Field}
\acrodef{mmap}[MMAP]{Marginal Maximum a Posteriori}
\acrodef{em}[EM]{Expectation-Maximization}
\acrodef{sem}[SEM]{Stochastic Expectation-Maximization}
\acrodef{snr}[SNR]{Signal-to-Noise Ratio}
\acrodef{nr}[NR]{Newton-Raphson}
\acrodef{mmse}[MMSE]{Minimum Mean Squared error}
\acrodef{rqf}[RQF]{Reconstruction Quality Factor}
\acrodef{pb}[PB]{pseudo-Bayesian}
\acrodef{map}[MAP]{Maximum a Posteriori}
\acrodef{tv}[TV]{Total Variation}
\acrodef{st}[s.t.]{such that}

\newcommand{\ee}{\,\ensuremath{\mathbf{e}}}

\newcommand{\nm}[0]{(n,m)}

\title{Instantaneous Frequency Estimation In Multi-Component Signals Using Stochastic EM Algorithm}

\author{Quentin Legros, Dominique Fourer, Sylvain Meignen, Marcelo A. Colominas}
  
\begin{document}

\maketitle

\begin{abstract}
 This paper addresses the problem of estimating the modes of an observed non-stationary mixture signal in the presence of an arbitrary distributed noise.
A novel Bayesian model is introduced to estimate the model parameters from the spectrogram of the observed signal, by resorting to the stochastic version of the EM algorithm to avoid the computationally expensive joint parameters estimation from the posterior distribution. 
The proposed method is assessed through comparative experiments with state-of-the-art methods.
The obtained results validate the proposed approach by highlighting an improvement of the modes estimation performance.
\end{abstract}

\section{Introduction}

In this work, we introduce a novel observation model for estimating the instantaneous frequency of modes of \ac{mcs} in the presence of an arbitrary distributed noise.
For this purpose, we use the signal spectrogram defined as the squared modulus of the \ac{stft}, and consider the 1D signals observed by selecting a given time instant of the spectrogram.
While performing parameters estimation using classical algorithms such as \cite{rilling2007one} is challenging in the presence of noise, more methods assuming the presence of spurious content \cite{laurent2020novel,legros:hal-03259013,brevdo2011synchrosqueezing,Carmona1997} achieve satisfactory estimation performance until average to low \ac{snr}. However these methods are generally made to deal with the presence of Gaussian noise and do not address the estimation problem in more complex scenarios (Poisson or gamma noise, mixture of noise).
Here, we develop a method in a Bayesian framework for estimating the modes \ac{if} where a mixture model is used to account for the presence of both the signal components and an arbitrary distributed noise. 
Prior models are associated with the parameters to model the available a priori knowledge.
An adapted estimation strategy is then formulated through a \ac{sem} algorithm \cite{EM_Book,celeux1996stochastic} to avoid intractable joint parameters estimation from the posterior distribution and to deal with the Markovian nature of the prior models.
This paper is organized as follows. In Section \ref{sec:model} we introduce the observation model and the prior distributions associated with the model parameters. We then discuss the estimation strategy in Section \ref{sec:strategy}, and comparatively evaluate the performance of the proposed method in Section~\ref{sec:results} with numerical results. Conclusions and future work are finally reported in Section~\ref{sec:conclusion}.


\section{Observation model} \label{sec:model}

Let $x$ be a mixture made of $K$ superimposed \ac{amfm} components expressed as:
\begin{equation}\label{eq:mcs}
 x(n) = \sum_{k=1}^{K} x^{k}(n)  \text{,\quad with } x^{k}(n) = a^{k}(n)\ee^{j\phi^{k}(n)},
\end{equation}
where $a^{k}(n)$ and $\phi^{k}(n)$ are respectively the time-varying amplitude and angular phase of the $k$th component at time $n$.
The discrete-time \ac{stft} of a signal $x$, using an analysis window $\theta$ s.t. $\theta(n)=\frac{1}{\sqrt{2\pi}L}\ee^{-\frac{n^{2}}{2L^{2}}}$, with $L$ being the time spread parameter, can be defined at each time instant $n\in [0,N-1]$ and each frequency bin $m\in [0,M-1]$, as:
\begin{equation}
F^{\theta}_x\nm = \displaystyle\sum_{l=-\infty}^{+\infty} x(l) \theta(n-l)^* \ee^{-j\frac{2\pi l m}{M}},
\end{equation}
with $z^*$ the complex conjugate of $z$. Let $\boldsymbol{S}=\{|F_x^\theta|^2\}_{n,m}$ be the spectrogram of $x$. 
We denote the $\mathds{R}^{M \times 1}$ spectrogram columns as $\boldsymbol{s}_{n}=[s_{n,0},\ldots,s_{n,M-1}]^{\top}$, with $\boldsymbol{S}=\{\boldsymbol{s}_{n}\}_{n=1}^N$.
In this work, we are interested in estimating the ridge positions $\boldsymbol{\hat{m}}_{n} = [\hat{m}_{n}^{1},\dots,\hat{m}_{n}^{K}]^{\top}$ associated with the \ac{if} $\boldsymbol{\phi}'_{n} = \{\frac{d\phi^{k}}{dn}(n)\}_{k=1}^K$, $\forall n \in [0,N-1]$.
For that purpose, we assume the following observation model: 
\begin{equation} \label{form:plik}
    p(s_{n,m}|\boldsymbol{w}_{n},\boldsymbol{\hat{m}}_{n})=\sum_{k=1}^{K}w_{n}^{k}g(m-\hat{m}_{n}^{k})+\frac{1}{M}\left(1- \sum_{k=1}^{K}w_{n}^{k} \right),
\end{equation}
where $g(m)=\frac{2\sqrt{\pi}L}{M} \ee^{-\left(\frac{2\pi m L}{M}\right)^{2}}$ is the normalized and discretized squared modulus of the Fourier transform of $\theta$, \ac{st} the integral of $g$ remains constant over the admissible values of $\boldsymbol{\hat{m}}_{n}$.
In \eqref{form:plik}, the weight $w_{n}^{k}$ represents the probability of each element of $\boldsymbol{s}_{n}$ to belong to the $k$th component \ac{st} $w_{n,k} = \frac{a_{n}^{k}}{\sum_{k=1}^{K}a_{n}^{k} + Mb_{n}}$, with $b_{n}$ the average noise amplitude at time $n$. Conversely, $(1- \sum_{k=1}^{K}w_{n}^{k})$ is the probability to observe noise in $\boldsymbol{s}_{n}$.
Note that the weights $\boldsymbol{w}_{n}=[w_n^1,\ldots,w_n^K]^{\top}$ are constrained to belong to $[0,1]^{K}$ \ac{st} $\sum_{k} w_{n}^{k}\leq 1$.
For further development, we set $\boldsymbol{W}=\{\boldsymbol{w}_{n}\}_{n=0}^{N-1}$ and  $\boldsymbol{\hat{\mathcal{M}}}=\{\boldsymbol{\hat{m}}_{n}\}_{n=0}^{N-1}$.
Assuming independence between each \ac{tf} instant conditioned on the value of $(\boldsymbol{w}_{n},\boldsymbol{\hat{m}}_{n})$, we express the joint likelihood function as
\begin{equation} \label{likelihood}
    p(\boldsymbol{S}|\boldsymbol{W},\boldsymbol{\hat{\mathcal{M}}}) = \prod_{n} \prod_{m} p(s_{n,m}|\boldsymbol{w}_{n},\boldsymbol{\hat{m}}_{n}).
\end{equation}
In order to complete the Bayesian model, prior distributions have to be assigned to the model parameters to account for the a priori available knowledge \cite{IATSENKO2016290}.
First, a weak uniform prior model is associated with the elements of $\boldsymbol{W}$.

\underline{Total Variation}
In the presence of strong noise (low \ac{snr}), the ridges in the \ac{tf} plane can be split or partially destroyed. It can thus be preferable for the \ac{if} estimates to not significantly move away from the estimation performed in the \ac{tf} area with strong local maxima. 
We thus define the following \ac{tv} \ac{mrf} prior model for $\boldsymbol{\hat{\mathcal{M}}}$ which preserves sharp edges \cite{chambolle2004algorithm,rudin1992nonlinear}
\begin{equation} 
\label{eq:prior_tv}
    p(\boldsymbol{\hat{\mathcal{M}}}|\epsilon) \propto \exp\left[-\epsilon\sum_{k=1}^K\| \boldsymbol{\hat{m}}_{k,:}^{\top}\|_{TV}\right],
\end{equation}
with $\boldsymbol{\hat{m}}_{k,:}$ the $k$th row of $\boldsymbol{\hat{\mathcal{M}}}$, $\epsilon$ a fixed user-defined hyper-parameter and $\|x\|_{TV}$ the sum of the absolute values of the partial derivatives of $x$.

\underline{Laplacian}
Another possible choice for regularizing $\boldsymbol{\hat{\mathcal{M}}}$ is to constrain the mean curvature of the estimated ridge, remaining to bound the \ac{if} second derivatives. This is performed using a \ac{mrf} Laplacian prior model \cite{Lap_operator,Lap_operator2} by setting a $\ell_{2}$-norm penalization on the curvature of $\boldsymbol{\hat{\mathcal{M}}}$ as
\begin{equation} 
\label{eq:prior_Lap}
    p(\boldsymbol{\hat{\mathcal{M}}}|\lambda) \propto \exp\left[-\frac{\lambda}{2}
    \sum_{k=1}^K\|\boldsymbol{L}_{a} \boldsymbol{\hat{m}}_{k,:}^{\top}\|_{2}^{2}\right],
\end{equation}
where $\boldsymbol{L}_{a}$ is the log-concave and differentiable Laplacian operator, \ac{st} $\|\boldsymbol{L}_{a}x\|_{2}^{2}$ is the sum of the squared partial derivatives of $x$. This operator controls the smoothness of the estimation. Similarly to the \ac{tv} prior model, we assume $\lambda$ to be a fixed hyperparameter.


\section{Estimation strategy}\label{sec:strategy}

For the sake of clarity, we omit in the sequel the priors related hyper-parameters $\epsilon$ and $\lambda$ from the equations.  
Using Bayes rule, the joint posterior distribution of $(\boldsymbol{w},\boldsymbol{\hat{\mathcal{M}}})$ can be approximated as 
\begin{equation} \label{Posterior}
    p(\boldsymbol{W},\boldsymbol{\hat{\mathcal{M}}}|\boldsymbol{S}) \propto  p(\boldsymbol{S}|\boldsymbol{W},\boldsymbol{\hat{\mathcal{M}}})p(\boldsymbol{\hat{\mathcal{M}}})p(\boldsymbol{W}).
\end{equation}

Estimating jointly $(\boldsymbol{W},\boldsymbol{\hat{\mathcal{M}}})$ is challenging due to the shape of the likelihood in Eq.~\eqref{likelihood} being multimodal with respect to $\boldsymbol{\hat{\mathcal{M}}}$. Thus, we propose to marginalize over the hidden parameter $\boldsymbol{\hat{\mathcal{M}}}$ to perform the estimation of $\boldsymbol{W}$ as the \ac{mmap} estimation as follow
\begin{equation} \label{eq:comp_w}
\widehat{\boldsymbol{W}}_{\text{MMAP}} =\underset{\boldsymbol{W}}{\textrm{argmax}}~~p(\boldsymbol{W}|\boldsymbol{S}).
\end{equation}
\ac{em}-based algorithms are particularly adapted to address this problem. Moreover, the shape of the model in Eq.~\eqref{form:plik} is well suited to apply such methods. It remains to compute $p(\boldsymbol{W}|\boldsymbol{S}) = \sum_{\boldsymbol{\hat{\mathcal{M}}}} p(\boldsymbol{W},\boldsymbol{\hat{\mathcal{M}}}|\boldsymbol{S})$, before solving Eq.~\eqref{eq:comp_w} in a second time.

\subsection{Estimation of mixture weights}
Given $\boldsymbol{W}^{(i)}$ the current estimation of $\boldsymbol{W}$ at iteration $i$, the E-step is given at each iteration by 
\begin{equation} \label{eq:estep}
        Q(\boldsymbol{W}|\boldsymbol{W}^{(i)}) = E_{\boldsymbol{\hat{\mathcal{M}}}|\boldsymbol{S},\boldsymbol{W}^{(i)}}\left[\log(p(\boldsymbol{S},\boldsymbol{\hat{\mathcal{M}}}|\boldsymbol{W}))\right].
\end{equation}
While classical \ac{em}-based algorithms are well adapted to solve problems involving hidden parameters, performing the M-step is computationally intractable due to the Markovian nature of the prior models introduced in Eq.~\eqref{eq:prior_tv}-\eqref{eq:prior_Lap}. We thus resort to the \ac{sem} \cite{EM_Book,celeux1996stochastic,9287414} \ac{st} $p(\boldsymbol{\hat{\mathcal{M}}}|\boldsymbol{W}^{(i)},\boldsymbol{S})$ in Eq.~\eqref{eq:estep} approximated using Markov chain Monte Carlo (MCMC) simulations. More precisely, we simulate $N_{s}$ samples $\{\boldsymbol{\bar{\mathcal{M}}}_{1},\ldots,\boldsymbol{\bar{\mathcal{M}}}_{N_{s}}\}$ from $p(\boldsymbol{\hat{\mathcal{M}}})$ using a 2-step Gibbs sampler as
\begin{equation} \label{eq:S-step}
    \boldsymbol{\bar{m}}_{n} = \underset{\boldsymbol{\hat{m}}_{n}}{\textrm{argmax}} \sum_{l=0 , l\neq n}^{N-1} p(\boldsymbol{\hat{m}}_{n}|\boldsymbol{\hat{m}}_{l}),
\end{equation}
with $\boldsymbol{\bar{\mathcal{M}}}=\{\boldsymbol{\bar{m}}_{n}\}_{n=0}^{N-1}$.
These samples are then used as an approximation of $p(\boldsymbol{\hat{\mathcal{M}}})$ and Bayes rule is applied to compute an approximate distribution $\tilde{p}(\boldsymbol{\hat{\mathcal{M}}}|\boldsymbol{W}^{(i)},\boldsymbol{S})$. We finally compute a current estimation $\boldsymbol{\widetilde{\mathcal{M}}}$ of $\boldsymbol{\hat{\mathcal{M}}}$ from $\tilde{p}(\boldsymbol{\hat{\mathcal{M}}}|\boldsymbol{W}^{(i)},\boldsymbol{S})$ using a sequential \ac{mmap} estimation which will be discussed in Section \ref{sec:estim_IF}. 
The convergence speed of the algorithm is increased by hot-starting the Gibbs sampler at each iteration using the previously generated samples.
The two main steps of the EM algorithm become 
\begin{equation}
\label{eq:modified_EM}
\begin{split}
\widehat{Q}(\boldsymbol{W}|\boldsymbol{W}^{(i)})&=\log p(\boldsymbol{S}|\boldsymbol{W},\boldsymbol{\widetilde{\mathcal{M}}}) + \log \left[p(\boldsymbol{\hat{\mathcal{M}}})p(\boldsymbol{W})\right],\\
\boldsymbol{W}^{(i+1)}&=\underset{\boldsymbol{W}}{\textrm{argmax}}~~ \widehat{Q}(\boldsymbol{W}|\boldsymbol{W}^{(i)}).
\end{split}
\end{equation}
The concavity of the likelihood in Eq.~\eqref{form:plik} with respect to $\boldsymbol{W}$ ensures that of $\widetilde{Q}(\boldsymbol{W}|\boldsymbol{W}^{(i)})$, allowing the use of convex optimization approaches to solve the M-step in Eq.~\ref{eq:modified_EM}. Maximization is thus performed using a Newton-Raphson second-order gradient ascent algorithm to update $\boldsymbol{W}$.
We set the same \ac{em} algorithm stopping criterion as in \cite{TCI_SWMSL}.

\subsection{Instantaneous frequency estimation}\label{sec:estim_IF}
Although having marginalized over the nuisance variable to estimate $\boldsymbol{W}$ allows to significantly reduce the computational cost of the whole estimation process, the approximate posterior in Eq.~\eqref{eq:S-step} remains non convex due to the presence of multiple components. Here, we estimate $\boldsymbol{\widetilde{\mathcal{M}}}$ by iteratively performing \ac{mmap} estimation from $p(\boldsymbol{\hat{\mathcal{M}}}|\boldsymbol{\widetilde{W}},\boldsymbol{S})$, before discarding the estimate neighborhood in the posterior distribution, until $K$ estimates have been computed.
Even though discarding $3$ times the standard deviation $\sigma_{d}=\sqrt{M/(\pi L)}$ of the data distribution would be a good choice (three-sigma rule of thumb), the presence of slightly frequency modulated components \cite{colominas2019} avoided proper discard of the information related to the last \ac{mmap} estimate. We thus consider a slightly broader window by considering $3\sigma_{d}+1$, rounded up.
Note that this choice depends on the frequency resolution of the \ac{stft}. 

\section{Results} \label{sec:results}

\begin{figure}[htb]
\centering
\includegraphics[trim={45pt 0pt 40pt 5pt}, clip, scale = 0.6]{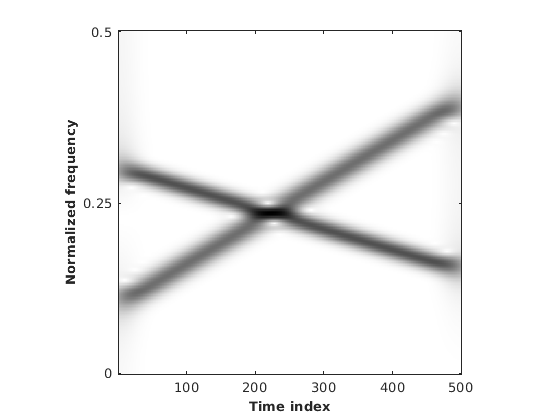}
\caption{Spectrogram of the analyzed multicomponent signals with overlapping components.}
\label{fig:spectrogram}
\end{figure}

In this section, we assess the \ac{if} estimation performance of the proposed approach on a \ac{mcs} in the presence of additive noise.
The \ac{mcs} depicted in Fig.~\ref{fig:spectrogram} is made of two linear chirps overlapping at time index $225$. 
For the experiments conducted in this section, we compute the \ac{stft} using the ASTRES toolbox \cite{fourer2017c}, with $N=500$ and $L=20$. 
Moreover, an additive white Gaussian noise controlled through a \ac{snr} varying from -20 to 20 dB, is added to the \ac{mcs} in order to model the presence of a spurious content.
For comparison purpose, we select the \ac{pb} approach proposed in \cite{legros:hal-03259013}, both the \textit{simple} and \textit{spline} \ac{rd} of \cite{laurent2020novel} and the Brevdo method \cite{brevdo2011synchrosqueezing}. For the latter we keep the same hyperparameter values than that presented in \cite{brevdo2011synchrosqueezing}.
For the proposed method, we set $\epsilon=10^{-3},\lambda=10^{-2}$ since it provides the best performances during our experiments.
Note however that those values have to be defined according to the \ac{tf} resolution.
For each method, we reconstruct the signals by selecting at each time instant in the \ac{tfr}, a neighborhood of $2\times 3 \sigma_{d} +1$ frequency values centered around the estimated \ac{if} of each component. This ensures most of the ridges information to be encapsulated in informative ribbons. A hard threshold is then applied on the \ac{tfr} and the \ac{tf} instants outside of the ribbons are set to zeros value. The inverse \ac{stft} is finally applied on the hard thresholded \ac{tfr} in order to compare the estimated signal to its ground truth. The estimation performance of the method is thus assessed using the \ac{rqf}:
$10 \log_{10}\left( \frac{||x||^2}{||x-\hat{x}||^2}\right)$
where $x$ (resp. $\hat{x}$) stands for the reference (resp. estimated) signal.\\
\begin{figure}[htb]
\centering
\includegraphics[trim={67pt 0pt 85pt 25pt}, clip, scale = 0.55]{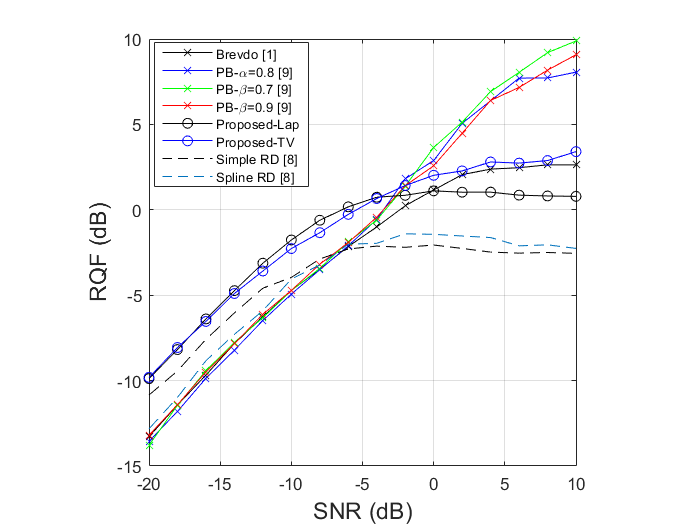}
\caption{\ac{rqf} of the first ridge with increasing frequency depicted in Fig.~\ref{fig:spectrogram} obtained with the competing methods (averaged over 50 realizations of noise) for a varying \ac{snr}.}
\label{fig:RQF1}
\end{figure}
\clearpage
\begin{figure}[htb]
\centering
\includegraphics[trim={67pt 0pt 85pt 25pt}, clip, scale = 0.55]{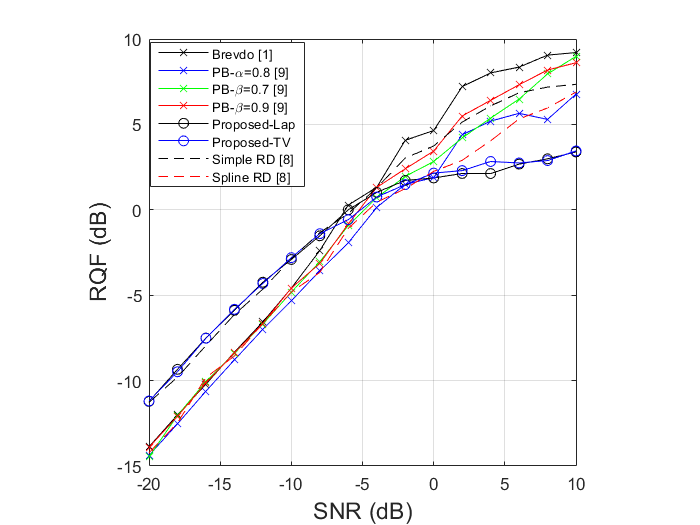}
\caption{\ac{rqf} of the second ridge with decreasing frequency depicted in Fig.~\ref{fig:spectrogram} obtained with the competing methods (averaged over 50 realizations of noise) for a varying \ac{snr}.}
\label{fig:RQF2}
\end{figure}

From Fig.~\ref{fig:RQF1} and Fig.~\ref{fig:RQF2}, we observe that the proposed approach using both prior models outperforms other at low \ac{snr}. The performance of the proposed method remains nonetheless similar to that of the \textit{spline} \ac{rd} \cite{laurent2020novel} in Fig.~\ref{fig:RQF2}.
Note that the experiment, and more particularly the reconstruction of the \ac{tfr} using a hard threshold, is not adapted to high \ac{snr} cases since we mostly discard informative content. The \ac{rqf} is however adapted for low \ac{snr} scenarios, even though the \ac{tf} content surrounding the estimated \ac{if} used to reconstruct the signals is also contaminated by noise.
Both the \ac{tv} and Laplacian prior models in the proposed approach provide similar performance, even though the \ac{tv} prior perform better at high \ac{snr} (see Fig.~\ref{fig:RQF2}). %

\section{Conclusion}\label{sec:conclusion}

In this work, we introduced a new observation model to perform estimation of the \ac{if} of a signal component in the presence of an arbitrary distributed noise.
The presence of spurious content is approximated as a uniform distribution to allow the modelling of any additive noise.
The model formulation is well suited for inference using EM algorithms, allowing to reduce the problem complexity and to estimate the mixture weights with a low computational time. The stochastic approach significantly lightens the computation of the E-step when using Markovian prior models.
A sequential \ac{mmap} estimation is performed to account for the multimodal nature of the \ac{if} posterior distribution, even in scenarios involving overlapping ridges.
The results demonstrate the ability of the proposed inference method to outperform the competing approaches in the low \ac{snr} regime. 
Future work include an estimation of the modulation rate \cite{colominas2019} by extending, for instance, the method into a generalized EM \cite{EM_Book}. A more general estimation process accounting for the estimation of the hyperparameters is currently under investigation.

\small 

\end{document}